\documentclass[12pt,a4paper]{article}
\usepackage{amstext}
\usepackage{graphics}
\newcommand{\nn}{\noindent}

\newcommand{\hed}[1]{\noindent{\bf #1 \\}}
\newcommand{\bq}{\begin{equation}}
\newcommand{\eq}{\end{equation}}
\newcommand{\ba}{\begin{eqnarray}}
\newcommand{\ea}{\end{eqnarray}}
 
\begin{document}
\thispagestyle{empty}
\onecolumn
\begin{flushleft}
Interner Bericht 
\\
DESY-Zeuthen 96--05
\\
LMU--03/96
\\
Freiburg--THEP 96/5
\\
M\"arz 1996
\\
hep-ph/9603438
\end{flushleft}
\vspace*{.50cm}

\begin{center}

\LARGE
{{\tt{GENTLE/4fan}
\vspace*{.50cm}
}
\\
\Large
 A package of Fortran programs for the description 
\\
of $e^+ e^-$
 annihilation into four fermions$^{\dag}$
}

\vspace{1.0cm}

\nn
\large
D. Bardin$^{1,2}$\footnote{Supported by European Union with grant
  INTAS--93--744.} ,
~~~D. Lehner$^3$,
~~~A. Leike$^4$,
~~~T.~Riemann$^1$
\\

\end{center}

\vspace{0.7cm}
\noindent
\normalsize
{\small \it 
$^1$DESY -- Zeuthen,
          Platanenallee 6, D-15738 Zeuthen, Germany

\vspace{1mm}
\noindent
$^2$Bogoliubov Laboratory for Theoretical Physics, JINR,
    ul. Joliot-Curie 6, RU-141980 Dubna, 
\\
~Russia

\vspace{1mm}
\noindent
$^3$Fakult\"at f\"ur Physik, Albert-Ludwigs-Universit\"at,
Hermann-Herder-Str. 3, D-79104 Freiburg, 
\\
Germany

\vspace{1mm}
\noindent
$^4$Sektion Physik, Ludwig-Maximilians-Universit\"at,
Theresienstr. 37, D-80333 M\"unchen, 
\\
~Germany
} 

\vspace{\fill}
\vfill
\thispagestyle{empty}
{\large
\centerline{Abstract}
}
\small
\nn
We describe the program package {\tt gentle/4fan},
which 
consists of the Fortran codes {\tt gentle\_4fan.f}, {\tt 4fan.f},
and {\tt   gentle\_nc\_qed.f} and is 
devoted to the description of $e^+e^-$
annihilation into four fermions.
The codes are based on the semi-analytical approach.
Initial state QED corrections are taken into account.
The program versions, which are described here were used in the 1995
workshop `Physics at LEP~2'.  
\normalsize
\vfill

\bigskip
\vfill
\nn
{\it
$^{\dag}$
Contributions to the working groups on
`Event Generators for WW Physics' 
and 
`Event Generators for Discovery Physics' 
of the 1995 Workshop on
Physics at LEP~2, to appear in the proceedings (G. Altarelli,
T. Sj\"ostrand and F. Zwirner (eds.), CERN 96--01)
}
\normalsize
\vfill \eject

\clearpage
\section{Introduction}
This note collects two slightly updated contributions to the 
working groups on 
`Event Generators for WW Physics'~\cite{gentle} 
and on 
`Event Generators for Discovery Physics'~\cite{gentle_4fan} 
of the Workshop on `Physics at LEP~2' held at
CERN in 1995. 
The underlying formulae may be found
in references~\cite{gentle_nunicc}--\cite{gentle_nuninc}. 
In this series of papers we advocate a semi-analytical approach to the
reaction
\bq\nonumber
e^+e^-\rightarrow f_1\bar f_1 f_2\bar f_2 (n\gamma).
\eq
The approach is characterized by the attempt to perform at least part
of the eight-dimensional phase space integration analytically. 
As a result, the numerical integrations are smoothened and often
faster than in a completely numerical approach.
Further, the numerical accuracy is well under control.
The analytical integrations are performed over (part of) the angular
variables, while the boson invariant masses are left for
numerical integration.

Four-fermion production processes may be very
complex and proceed via a large variety of intermediate states.
Doubly resonant processes are usually calles
`basic processes'.
The others are called background contributions.
We repeat the classification of reference~\cite{wwteup},
with the number of Feynman diagrams in the charged current (\emph{CC})
classes are shown in table~\ref{tab1} 
and for the final states corresponding to the neutral current
(\emph{NC}) classes in table~\ref{tab2}.

\begin{table}[ht]
\begin{center}
\begin{tabular}{|c|c|c|c|c|c|}
\hline
             &
\raisebox{0.pt}[2.5ex][0.0ex]{${\bar d} u$}
& ${\bar s} c$ & ${\bar e} \nu_{e}$ &
              ${\bar \mu} \nu_{\mu}$ & ${\bar \tau} \nu_{\tau}$   \\
\hline
$d {\bar u}$            &{\it  43}& {\bf 11} &  20 & {\bf 10} & {\bf 10} \\
\hline
$e {\bar \nu}_{e}$      &  20 &  20 &{\it 56}&  18 &  18 \\
\hline
 $\mu {\bar \nu}_{\mu}$ & {\bf 10} & {\bf 10} &  18 & {\it 19} & {\bf 9}  \\
\hline
\end{tabular}
\caption
{\it
Number of Feynman diagrams for class CC final states.
\label{tab1}
}
\end{center}
\end{table}
%
\begin{table}[ht]
\begin{center}
 \begin{tabular}{|c|c|c|c|c|c|c|}
\hline
&
\raisebox{0.pt}[2.5ex][0.0ex]{${\bar d} d$}
&${\bar u} u$
&${\bar e} e$
&${\bar \mu} \mu$
&${\bar \nu}_{e} \nu_{e}$
&${\bar \nu}_{\mu} \nu_{\mu}$
\\
\hline
\raisebox{0.pt}[2.5ex][0.0ex]{${\bar d} d$}
 & {\tt 4$\cdot $16} & {\it 43} & {48}
             & {\bf 24} & 21 & {\bf 10} \\
\hline
\raisebox{0.pt}[2.5ex][0.0ex]
{${\bar s} s, {\bar b} b$} & {\bf 32} & {\it 43} & {48}
             & {\bf 24} & {21} & {\bf 10} \\
\hline
${\bar u} u$ & {\it 43} & {\tt 4$\cdot$16} & {48}
             & {\bf 24} & {21} & {\bf 10} \\
\hline
${\bar e} e$ &{48} &{48} & \textsf{4$\cdot$36} &{48}
& {\it 56} & {20}
\\
\hline
${\bar \mu} \mu$  & {\bf 24} & {\bf 24} & {48} & {\tt 4$\cdot$12}
                  & {19} & {\it 19}         \\
\hline
${\bar \tau} \tau$& {\bf 24} & {\bf 24} & {48} & {\bf 24}
                  & {19} & {\bf 10}         \\
\hline
${\bar \nu}_e \nu_{e}$  & {21} & {21} & {\it 56} & {19}
                  & \textsf{4$\cdot$9} & {12}                   \\
\hline
${\bar \nu}_{\mu} \nu_{\mu}$ & {\bf 10} & {\bf 10} & {20}
             & {\it 19} & {12} & {\tt 4$\cdot$3}  \\
\hline
${\bar \nu}_{\tau} \nu_{\tau}$ & {\bf 10} & {\bf 10} & {20}
             & {\bf 10} & {12} & {\bf 6}  \\
\hline
\end{tabular}
\caption[]
{\it
Number of Feynman diagrams for class NC final states.
\label{tab2}
}
\end{center}
\end{table}

\section{{\tt GENTLE/4fan}}
%
\leftline{\bf Authors:}
 
\begin{tabular}{ll}
D. Bardin$^a$    & {\tt bardin@ifh.de} \\
M. Bilenky$^a$   & {\tt bilenky@ifh.de} \\
D. Lehner$^b$    & {\tt lehner@phyv4.physik.uni-freiburg.de} \\
A. Leike$^a$     & {\tt leike@graviton.hep.physik.uni-muenchen.de} \\
A. Olchevski$^a$ & {\tt OLSHEVSK@VXCERN.CERN.CH} \\
T. Riemann$^a$   & {\tt riemann@ifh.de}
\end{tabular}

\indent $ ^a$ {\sc Fortran} code {\tt gentle\_4fan.f} \\
\indent $ ^b$ {\sc Fortran} code {\tt gentle\_nc\_qed.f}
 
\bigskip

\noindent{\bf Description of the package}
 
\noindent
The {\tt GENTLE/4fan} package is designed to compute selected total
four-fermion production cross-sections and final-state fermion pair
invariant mass distributions for charged current (\emph{CC}) and
neutral current (\emph{NC}) mediated processes within the Standard
Model (SM).
For the \emph{CC03} subprocess, the $W$ production angular distribution
is also accessible.
In the \emph{NC} case, SM Higgs Production is included.
The phase space integration is carried out by a semi-analytical
technique, which is described below.
The {\tt GENTLE/4fan} package is written in {\tt Fortran}.
It consists of two branches.
The basic branch {\tt gentle\_4fan.f} contains all features of the
package but complete initial-state radiation (ISR) to \emph{NC} processes.
The subroutine {\tt FOURFAN}, which is called by {\tt gentle\_4fan.f}
performs 
the computation of \emph{NC} cross-sections and is described
in the next section.
The independent branch {\tt gentle\_nc\_qed.f} includes complete
ISR to \emph{NC02} and \emph{NC08} and will be merged into
{\tt gentle\_4fan.f}.
 
\noindent{\bf Program features:}
\begin{enumerate}
\item\hed{Method of integration:}
The package is a {\em semi-analytical}~one. Without (with) ISR,
the phase space is parame\-trized by five (seven) angular variables
and the final-state fermion pair invariant masses (plus the reduced
center of mass energy squared).
All angular phase space variables are integrated analytically.
The resulting formulae are input to the package.
Invariant masses are subsequently integrated numerically with a
self-adaptive Simpson algorithm.
Optionally, for the \emph{CC03} subprocess, the W production angle may
also be numerically integrated.
The method is numerically stable and, in most cases, very fast.
 
\item\hed{Possible final states:}
The package may treat all four-fermion final states which do not
contain identical particles, electrons, or electron neutrinos. This
means that the package accesses all final states that are described by
{\em annihilation} and {\em conversion} type Feynman diagrams:

\newpage

\begin{enumerate}
  \item[(1)] \emph{CC03} (with complete ISR)~\cite{gentle_nunicc}
  \item[(2)] \emph{NC02}, \emph{NC08} (with complete ISR)~\cite{gentle_nuninc}
  \item[(3)] \emph{CC9}, \emph{CC10}, \emph{CC11}~\cite{gentle_unicc11}
  \item[(4)] \emph{NC06}, \emph{NC10}, \emph{NC24},
    \emph{NC32}~\cite{gentle_nc24}
  \item[(5)] \emph{NC} + Higgs~\cite{gentle_nc24h}
\end{enumerate}
Via flags, cross-sections for subsets of Feynman diagrams may be
extracted.
 
\item\hed{Cuts}
Cuts may be imposed on invariant masses of fermion pairs and on the
invariant mass of the final-state four-fermion system.
Using the structure function approach in {\tt gentle\_4fan.f}, cuts on
the electron/positron momentum fraction can be imposed.
For the \emph{CC03} subprocess, cuts on the $W$ production angle are
enabled.
 
\item\hed{Initial state radiation}
ISR is implemented into the package.
{\em Universal}~ISR is present for all processes~\cite{gentle_unicc11}.
In addition, the package includes complete, i.e. {\em universal}~and
{\em non-universal}~ISR for the \emph{CC03}, \emph{NC02}, and \emph{NC08}
processes~\cite{gentle_nunicc,gentle_nuninc}.
{\em Non-universal}~ISR does not contribute to {\em annihilation}
diagrams.
It may be argued that {\em non-universal}~ISR is very small,
${\cal O}(10^{-3})$, for {\em conversion}-{\em annihilation}
interferences.
The speed of the package is reduced, if {\em non-universal}~ISR
is included, due to its complex analytical structure.
 
\item\hed{Final state radiation}
Final state radiation is not implemented.
 
\item\hed{Treatment of final state decays}
Final state decays are not accounted for.
 
\item\hed{Treatment of the Coulomb Singularity}
The Coulomb singularity is included according to reference~\cite{BBD}.
 
\item\hed{Treatment of the Anomalous Couplings}
Anomalous couplings are not included.
 
\item\hed{Treatment of masses}
In general, final-state masses are neglected in the matrix
elements. Where needed, however, masses are retained in the phase
space. In addition, masses of heavy particles coupling to the Higgs
boson are taken into account where appropriate; see section 3.
 
\item\hed{Hadronization}
No interface to hadronization is foreseen.
 
\end{enumerate}
 
\newpage

\noindent{\bf Input parameters}
 
\noindent
All input parameters are set inside the {\tt Fortran} code.
{\tt gentle\_4fan.f} uses the following flags, set in the subroutine
{\tt WWIN00}:
 
\begin{tabular}{ll}
 
 {\tt IBCKGR}: & \emph{CC03} case ({\tt IBCKGR}=0) or \emph{CC11} case
                 ({\tt IBCKGR}=1) \\
 {\tt IBORNF}: & Tree level ({\tt IBORNF}=0) or ISR corrected
                 ({\tt IBORNF}=1) quantities \\
 {\tt ICHNNL}: & \emph{CC03} ({\tt {ICHNNL}}=0), \emph{CC11} with
                 specific final state $\left[ l_1\nu_1 l_2\nu_2
                 ({\tt {ICHNNL}}=1), l\nu q{\bar q} \right. $ \\
               & $\left. ({\tt {ICHNNL}}=2,3), \;\;
                 q_1{\bar q}_1 q_2{\bar q}_2~
                 ({\tt {ICHNNL}}=4)\right]$, and
                 inclusive \emph{CC11} ({\tt ICHNNL}=5)  \\
 {\tt ICOLMB}: & Inclusion of Coulomb singularity ({\tt
                 ICOLMB}=1,...,5) or not ({\tt ICOLMB}=0) \\
               & Recommended value: {\tt ICOLMB}=2 \\
 {\tt ICONVL}: & Flux function ({\tt ICONVL}=0) or structure
                 function approach ({\tt ICONVL}=1) for  \\
               & ISR. Recommended value: {\tt ICONVL}=0  \\
 {\tt IGAMZS}: & Constant $Z$ width ({\tt IGAMZS}=0) or $s$-dependent
                 $Z$ width ({\tt IGAMZS}=1) \\
 {\tt IINPT}:  & Input for tuned comparison~\cite{bakl} ({\tt
                 IINPT}=0) or 
                 preferred input ({\tt IINPT}=1) \\
 {\tt IIQCD}:  & Naive inclusive QCD corrections are included
                 ({\tt IIQCD}=1) or not ({\tt IIQCD}=0) \\
 {\tt IMMIM}:  & Minimum number of a moment requested by
                 {\tt IREGIM}\\
 {\tt IMMAX}:  & Maximum number of a moment requested by
                 {\tt IREGIM}\\
 {\tt IONSHL}: & On-shell ({\tt IONSHL}=0) or off-shell heavy bosons
                 ({\tt IONSHL}=1) \\
 {\tt IPROC}~: & \emph{CC} case ({\tt IPROC}=1) or \emph{NC} case
                 ({\tt IPROC}=2, call to {\tt FOURFAN} is initialized)
                  \\
 {\tt IQEDHS}: & Determination of the {\em universal} ISR radiator: \\
               & \indent ${\cal O}(\alpha)$ exponentiated
                 ({\tt IQEDHS}=--1,0); \\
               & \indent ${\cal O}(\alpha)$ exponentiated plus
                 different ${\cal O}(\alpha^2)$ contributions ({\tt
                 IQEDHS}=1,...,4) \\
               & Recommended value: {\tt IQEDHS}=3 
\\
 {\tt IREGIM}: & Calculation of the total cross-section ({\tt
                 IREGIM}=0), the moments of the radi- \\
               & ative loss of final-state four-fermion invariant mass
                 ({\tt IREGIM}=1), the moments \\
               & of the radiative energy loss ({\tt IREGIM}=2), the
                 moments of the $W$ mass shift \\
               & $\left(\sqrt{s_+} \!+\! \sqrt{s_-} \!-\! 2
                 M_W\right)$ ({\tt IREGIM}=3), and the first moments of
                 $\cos\left( n\theta_W \right)$, \\
               & $n=1,...,4$ ({\tt IREGIM}=4) \\
 {\tt IRMAX}~: & Maximum value of {\tt IREGIM} \\
 {\tt IRSTP}~: & Step in a DO loop over {\tt IREGIM} \\
 {\tt ITVIRT}: & {\em Non-universal} virtual ISR included
                 ({\tt ITVIRT}=1) or not ({\tt ITVIRT}=0) \\
 {\tt ITBREM}: & {\em Non-universal} bremsstrahlung included
                 ({\tt ITBREM}=1) or not ({\tt ITBREM}=0) \\
 {\tt IZERO}~: & See equation (4.5) of~\cite{gentle_unicc11}.
                 Recommended value: {\tt IZERO}=1 \\
 {\tt IZETTA}: & See equation (4.21) of~\cite{gentle_unicc11}.
                 Recommended value: {\tt IZETTA}=1
\end{tabular}
 
\noindent
In the {\tt gentle\_nc\_qed.f} branch, only the flags {\tt IBORNF, IONSHL,
  ITVIRT, ITBREM} are used. The additional flag {\tt IBOSON} in
{\tt gentle\_nc\_qed.f} distinguishes between the \emph{NC02} and the 
\emph{NC08} processes.
 
\noindent
The center of mass energy squared is chosen by setting the variable
{\tt IREG} and the parameters {\tt ISMAXA} or {\tt ISMAXB} in the main
program.
The following input may be changed by the user:
\begin{center}
\begin{tabular}{rcccl}
     {\tt GFER} & = &  $G_\mu$ & = &
        1.16639 $\times 10^{-5}$ GeV$^{-2}$, the Fermi coupling constant \\
     {\tt ALPW} & = &  $\alpha(2 M_W)$ & = &
        1/128.07, the running fine structure constant at $2\,M_W$ \\
     {\tt AME}  & = &  $m_e$ & = &
        0.51099906 $\times 10^{-3}$ GeV, the electron mass \\
     {\tt AMZ}  & = &  $M_Z$ & = & 91.1888 GeV, the $Z$ mass, \\
     {\tt AMW}  & = &  $M_W$ & = & 80.230 GeV, the $W$ mass \\
     {\tt GAMZ} & = &  $\Gamma_Z$ & = & 2.4974 GeV, the $Z$ width \\
     {\tt ALPHS}& = &  $\alpha_{_S}(2M_W)$ & = & $0.12$ 
\end{tabular}
\end{center}
 
\noindent{\bf Output}
 
\noindent
The following derived quantities are computed in {\tt gentle\_4fan.f}
and printed in the output:
\begin{eqnarray}
  {\tt GAMW} & = & \Gamma_W \; = \;
    \frac{9}{6\sqrt{2}\pi} \, G_{\mu}M_W^3
    \left( 1+\frac{2 \alpha_{_S}(2M_W) }{3\pi} \right)
    \nonumber \\
  {\tt SIN2W} & = & \sin^2 \theta_W \; = \; 1 - M_W^2/M_Z^2
    \nonumber \\
  {\tt GAE} & = & - \frac{e}{4s_Wc_W} \; = \;
    - \frac{\sqrt{4\pi\alpha(2M_W)}}{4s_Wc_W}
    \nonumber \\
  {\tt GVE} & = & {\tt GAE} \cdot (1-4s_W^2)
    \nonumber \\
  {\tt GWF} & = & \frac{g}{2\sqrt{2}} = - {\tt GAE} \cdot \sqrt{2} c_W
    \nonumber \\
  {\tt |GWWG|} & = & \sqrt{4\pi\alpha(2M_W)}
    \nonumber \\
  {\tt |GWWZ|} & = & {\tt |GWWZ|} \cdot \frac{c_W}{s_W}
    \nonumber
\end{eqnarray}
{\tt GVE} and {\tt GAE} are the electron vector and axial vector
couplings, {\tt GWF} is the fermion-$W$ coupling, and {\tt |GWWG|} and
{\tt |GWWZ|} are the trilinear gauge boson couplings for the photon
and the $Z$ respectively.
Further the output repeats the flag settings.
After the cross-section calculation, the following output is printed:
\begin{eqnarray}
  {\tt SQS}  & = & \sqrt{s}
    \nonumber \\
  {\tt XSEC0} & = & \sigma_{\rm tot}(s)  \hspace{.5cm} {\rm in~nanobarns}
\end{eqnarray}
In addition, the calculated {\tt MOMENTS} are printed. In the first
column {\tt IREGIM} is printed.
The second column is arranged in blocks of three lines each.
The first line contains the integer $n$.
The second line contains the $n^{th}$ moment of the physical quantity
indicated by {\tt IREGIM}.
The third line contains the dimensionless $n^{th}$ moment obtained
through division of the $n^{th}$ moment by the proper power of
$\sqrt{s}/2$.
 
\noindent
Although variable names are slightly different,
{\tt gentle\_nc\_qed.f} uses the same derived quantities as
{\tt gentle\_4fan.f}.
For one run, {\tt gentle\_nc\_qed.f} outputs the used flag values
together with the fermion code numbers {\tt IFERM1/IFERM2}, the color
factors {\tt RNCOU1/RNCOU2}, the masses {\tt AM1/AM2}, and the
invariant pair mass cuts {\tt CUTM12,CUTM34} for the final-state
fermion pairs.
In addition, the lower cut {\tt CUTXPR} on the ratio of the
four-fermion invariant mass squared over the center of mass energy
squared, $s'/s$\ is output.
The main output, however, is an array of center of mass energies and
the corresponding total cross-sections.
 
\noindent{\bf Availability}                         \\
The codes and this description are available from the authors upon
E-Mail request or via WWW 
 
\begin{tabular}{lcl}
  \hspace*{.5cm}
  {\tt gentle\_4fan.f} & from & {\tt http://www.ifh.de/}
        $\tilde{ }$ {\tt bardin/gentle\_4fan.uu} 
\\ \hspace*{.5cm}
  {\tt gentle\_nc\_qed.f} & from & {\tt http://www.ifh.de/}
        $\tilde{ }$ {\tt lehner/gentle\_nc\_qed.uu}
\\ \hspace*{.5cm}
  {DESY-Zeuthen 96--05} & from & {\tt http://www.ifh.de/theory/publist.html}
\end{tabular}
%
 
\clearpage
%

\section{{\tt 4fan version 1.3}}
\leftline{\bf Authors:}
 
\begin{tabular}{ll}
D. Bardin    & {\tt bardin@ifh.de} \\
A. Leike     & {\tt leike@graviton.hep.physik.uni-muenchen.de} \\
T. Riemann   & {\tt riemann@ifh.de}
\end{tabular}

\noindent{\bf Description of the package}
 
\noindent
{\tt 4fan} is a semi-analytical program which calculates the process
\bq\nonumber
e^+e^-\rightarrow f_1\bar f_1 f_2\bar f_2,
\eq
where the three involved fermions $e,f_1$ and $f_2$ must be in different 
electroweak multiplets (the {\it NC32} class)
\cite{gentle_nc24}; see table~2. 
Optionally, Standard Model Higgs production can be
included~\cite{gentle_nc24h}.
For calculations at the Born level, {\tt 4fan} can be used as a
stand-alone program. 
For the calculation of cross-sections including initial state radiation, the
initial state radiation environment of the code {\tt gentle\_4fan.f}
must be used which calls {\tt FOURFAN} as
a subroutine.
For the description of {\tt gentle\_4fan.f} we refer to section~2.
In the following, we describe the stand-alone program {\tt 4fan.f}:

Six of the eight integrations of the four particle phase space 
were done 
analytically. The two remaining integrations over 
$s_1=[p(f_1)+p(\bar f_1)]^2$ and $s_2=[p(f_2)+p(\bar f_2)]^2$
are performed numerically allowing the inclusion of cuts for these variables.

Finite mass effects are taken into account using the following 
approximations: 
\begin{itemize}
\item The phase space is treated exactly. 
\item In the Higgs contributions and the conversion diagrams
       $e^+e^-\rightarrow (\gamma\gamma)\rightarrow f_1\bar f_1 f_2\bar f_2$,
      the masses are treated up to order $O[m^2(f_i)/s_i]$.
\item Fermion masses are treated identically in traces and Higgs couplings.
\item The Higgs width is calculated including the decays into $b$-, $c$- 
      and $\tau$-pairs. 
\end{itemize}
The numbers quoted in the tables of the report~\cite{bakl} 
are produced for zero fermion masses except in the Higgs couplings.
The Higgs propagator is always connected with $s_2$ by convention.

The initialization routine {\tt BBMMIN} contains the input from the 
Particle Data Group~\cite{pdb}.
In the subroutine {\tt DSDSHSZ}, the interferences between the three main
subsets of the diagrams of the {\it NC32} class are calculated as well
as those with the Higgs signal diagram.
Their sum yields the double differential cross-section.
Single interferences between these subsets are not printed.

The numerical integration is done by a twofold application of a 
one-dimensional self-adaptive Simpson algorithm with control
over the relative and the absolute error.
The singularities due to resonant vector boson propagators are
eliminated by transformations of the integration variables. 
To avoid numerical instabilities,
the kinematical functions resulting from the six-fold analytical integration
are replaced by Taylor expansions near the borders of the phase space.
The shortest calculation time is achieved by a choice of the
required absolute and relative errors in such a way that
they give approximately equal contributions to the error of the output.

The calculation time of a Born cross-section is several
seconds on an HP workstation,
depending on the required accuracy and the cuts on $s_1$ and $s_2$.
Ten times higher accuracy needs approximately two times longer
calculational time.

Input and output are transferred through the arguments of the
subroutine only. 

Usage of the program:\\
{\tt\hspace{6ex} 
CALL FOURFAN(EPS,ABSE,IF1,IF2,S,S1MIN,S1MAX,S2MIN,S2MAX,AMH,IOUT,OUT)}
\vspace{0.3cm}\\ 
\nn{\bf Input:}\hfill\vspace{0.3cm}

\begin{tabular}{lll}
{\tt EPS,ABSE}:&& The required relative and absolute error. If one of
                \\ && 
                 the two criteria is fulfilled, the calculation
                stops. \\
{\tt IF1,IF2:}&& Integers specifying the two final fermion pairs      
                as in \\ && the Monte Carlo particle numbering scheme,
                see \\
                && Particle Data Group~\cite{pdb}, chapter 32.\\
{\tt S:} && The c.m. energy squared of the $e^+e^-$ pair.\\
{\tt S1MIN,S1MAX:} && The integration bounds of $s_1$.\\
{\tt S2MIN,S2MAX:} && The integration bounds of $s_2$.\\
{\tt AMH:} && The Higgs mass.\\
{\tt IOUT:}  && Integer, selecting the output. \\ &&
                Currently {\tt IOUT}=1, 2, 11 and 12 are implemented:\\
	& {\tt IOUT=1:} & Total cross-section $\sigma_t$ without Higgs.\\
	& {\tt IOUT=2:} & Differential cross-section 
                          ${\rm d}\sigma /{\rm d} s_2$ without Higgs.\\
	& {\tt IOUT=11, 12:} & The same as {\tt IOUT=1, 2} 
                             but {\it with} Higgs.
\end{tabular}\vspace{0.3cm}

The units of the input (if required) are GeV$^2$ or GeV.\vspace{0.3cm}

\nn {\bf Output:} {\tt OUT}\ \ \ 
Depends on the value of {\tt IOUT}.
The output is given in $fb$ or in $fb/$GeV.
\vspace{0.3cm}

\nn 
On HP workstations {\tt 4fan} must be compiled with the -K option. 

\noindent{\bf Availability}                         \\
The code and this description are available from the authors upon
E-Mail request or via FTP 
or WWW
 
\begin{tabular}{lcl}
  \hspace*{.5cm}
  {\tt 4fanv13.f} & from & {\tt ftp://gluon.hep.physik.uni-muenchen.de}
\\
  \hspace*{.5cm}
  {\tt 4fanv13.f} & from & {\tt http://www.ifh.de/theory/publist.html} 
\\ \hspace*{.5cm}
  {DESY-Zeuthen 96--05} & from & {\tt http://www.ifh.de/theory/publist.html}
\end{tabular}
%

\newpage

\section{Concluding remarks}
Not all of the final-state topologies of tables~1 and~2 have been treated by
our approach so far.
Those which have been treated belong to the classes {\it CC11} and
{\it NC32} and are printed in {\bf boldface} in the tables.

Presently we are studying the semi-leptonic and leptonic processes of
the classes {\it CC20, NC48, NC21}, which are
printed in the tables in roman~\cite{BaBiLeRi}.

For a complete treatment of the {\it NC32} class, the four jet
production from $e^+e^-$ annihilation into $q{\bar q}gg$ has to be
covered in addition to $4f$ production~\cite{JaLe}.

Besides a verification of the Standard Model predictions for $WW$ and
$ZZ$ pair production and in Higgs boson searches
via $ZH$ production,
one is also interested in searches for anomalous triple gauge boson
couplings.
A description of {\it CC03, CC11} cross-sections with anomalous
couplings is under development~\cite{BiRi}.    

For a detailed study of the underlying physics, angular
distributions are extremely helpful.
Although in the semi-analytical approach one has a limited flexibility
concerning distributions and cuts, we made several attempts to
calculate some of them~\cite{gentle_unicc11,BiRi,Le}.


\bigskip


 
\begin{thebibliography}{99}
\bibitem{gentle}
D. Bardin, D. Lehner, A. Leike and A. Olchevski,
contribution to the  report of the Working Group {\em Event Generators
for WW Physics}, D. Bardin and R. Kleiss (conveners),
in: 
G. Altarelli,
T. Sj\"ostrand and F. Zwirner (eds.), Proceedings of the Workshop
{\em Physics at LEP~2}, CERN 96--01, to appear.
\bibitem{gentle_4fan}
A. Leike and T. Riemann,
contribution to the  report of the Working Group {\em Event Generators
for Discovery Physics}, 
M.L. Mangano and G. Ridolfi (conveners), 
in: 
G. Altarelli,
T. Sj\"ostrand and F. Zwirner (eds.), Proceedings of the Workshop
{\em Physics at LEP~2}, CERN 96--01, to appear.
\bibitem{gentle_nunicc}
D. Bardin, M. Bilenky, A.~Olchevski and T.~Riemann,
{\em Off Shell $W$ Pair Production in $e^+e^-$ Annihilation:
Initial State Radiation},
\textit{Phys.\ Lett.} {\bf B308} (1993) 403; 
E: {\it Phys. Lett.} {\bf B357} (1995) 725;
complete revised version: hep-ph/9507277.  
\bibitem{BBD}
D.~Bardin, W.~Beenakker and A.~Denner,
{\em The Coulomb Singularity in Off Shell $W$ Pair
Production},
\textit{ Phys.~Lett.}~{\bf B317} (1993) 213.
\bibitem{wwteup}
D.~Bardin, M.~Bi\-len\-ky, D.~Leh\-ner, A.~Ol\-chev\-ski
and T.~Riemann,
{\em Semianalytical Approach to Four Fermion Production
in $e^+e^-$ Annihilation},
in:
T.~Riemann and J.~Bl\"umlein (eds.),
Proc. of the Zeuthen Workshop {\em Elementary Particle Theory --
Physics at LEP200 and Beyond}, Teupitz, Germany,
April 10--15, 1994,
{\it Nucl.\ Phys.}  (Proc.\ Suppl.) {\bf 37B} (1994) p. 148
[hep-ph/9406340]. 
\bibitem{gentle_nc24}
D. Bardin, A. Leike and T. Riemann,
{\em The Process $e^+e^- \to l{\bar l} Q {\bar Q}$ at
LEP and NLC},
\textit{Phys.\ Lett.} {\bf B344} (1995) 383.
\bibitem{gentle_nc24h}
D. Bardin, A. Leike and T. Riemann,
{\em Higgs Production in $e^+e^- \to l{\bar l} Q {\bar Q}$ at
LEP and NLC},
{\it Phys.\ Lett.} {\bf B353} (1995) 513.
\bibitem{gentle_unicc11}
D.~Bardin and T.~Riemann,
{\em Off Shell $W$ Pair Production in $e^+e^-$ Annihilation: the
CC11 Process},
preprint DESY~95--167 (1995), to appear in \textit{Nucl.\ Phys.}
\textbf{B} [hep-ph/9509341].
\bibitem{gentle_nuninc}
D. Bardin, D. Lehner and T.~Riemann,
{\em Complete Initial State QED Corrections to Off Shell
Gauge Boson Pair Production in $e^+e^-$ Annihilation},
preprint DESY 96--028 (1996) [hep-ph/9602409];
\\
D. Lehner,
{\em Initial State Radiative Corrections to $Z$ Pair Production in
  $e^+e^-$ Annihilation - The Semi-Analytical Approach},
Ph.D. thesis, Humboldt-Universit\"at zu Berlin (1995),
DESY-Zeuthen 95-07, [hep-ph/9512301],
available from \verb+http://www.ifh.de/~lehner+.
\bibitem{bakl}
D. Bardin and R. Kleiss et al.,
report of the Working Group {\em Event Generators
for WW Physics}, in: 
G. Altarelli,
T. Sj\"ostrand and F. Zwirner (eds.), Proceedings of the Workshop
{\em Physics at LEP~2}, CERN 96--01, to appear.
\bibitem{pdb}
Particle Data Group (L. Montanet et al.) {\em Review of Particle
Properties}, 
\textit{Phys.\ Rev.} {\bf D40} (1994) 1173.
\bibitem{BaBiLeRi}
D. Bardin, J. Biebel, A. Leike and T. Riemann, under investigation.
\bibitem{JaLe}
M. Jack and A. Leike, in progress.
\bibitem{BiRi}
J. Biebel and T. Riemann, in progress; 
a preliminary Fortran program exists.
\bibitem{Le}
 A. Leike, in:
B. Kniehl (ed.),
Proceedings of the Workshop
{\em Perspectives for electroweak interactions in $e^+e^-$ collisions},
Ringberg, Germany, February 1995 (World Scientific, Singapore, 1995),
p.~121 [hep-ph/9504358].
\end{thebibliography}
\end{document}